# Development of a Fine Grating on ZnS for a Wideband Spectral Disperser in Characterizing Exoplanets using Space-borne Telescopes

Keigo Enya[1,a*], Takashi Sukegawa[2,b], Shigeru Sugiyama[2,c], Fumihiro Iijima[2,d], Naofumi Fujishiro[3,e], Yuji Ikeda[3,f], Tomohiro Yoshikawa[3,g], and Michihiro Takami[4,h]

[1]Institute of Space and Astronautical Science, Japan Aerospace Exploration Agency,
3-1-1 Yoshinodai, Chuo-ku, Sagamihara, Kanagawa 252-5210, Japan

[2]Canon Inc., 3-30-2 Shimomaruko, Ohta-ku, Tokyo 146-8501, Japan

[3]Koyama Astronomical Observatory, Kyoto Sangyo University,
Motoyama, Kamigamo, Kita-ku, Kyoto 603-8555, Japan

[4]Institute of Astronomy and Astrophysics, Academia Sinica,
P.O. Box 23-141, Taipei 10617, Taiwan, ROC

[a]enya@ir.isas.jaxa.jp, [b]sukegawa.takashi@canon.co.jp, [c]sugiyama.shigeru@canon.co.jp, [d]iijima.fumihiro@canon.co.jp, [e]ippaturo@nifty.com, [f]ikeda@photocoding.com, [g]tomohiro@kyoto-nijikoubou.com, [h]hiro@asiaa.sinica.edu.tw

**Abstract**

We present the fabrication and optical testing of a fine grating on a ZnS substrate to be used as a wideband infrared spectral disperser and for which the primary application is measurement of the composition of the atmospheres of transiting exoplanets using space-borne infrared astronomical telescopes. A grating with a blaze angle of 2.1 deg. and pitch of 166.667 μm was constructed on a roughly flat 10 mm × 10 mm substrate with a maximum thickness of 1 mm. To obtain high accuracy, the sample was fabricated on a ZnS monocrystal using a high performance processing machine at Canon Inc. The surface roughness measured with a microscope interferometer was 2.6 nm rms. The shape of the fabricated grating edges was examined with a scanning electron microscope. The diffraction efficiency was evaluated by optical experiments at $\lambda = 633$ nm, 980 nm, and 1550 nm, and compared with the efficiencies calculated using a Fourier Modal Method. The results showed that the differences between the diffraction efficiencies obtained from experiment and by calculation were between just 0.9 % and 2.4 %. We concluded that the quality of the fabricated ZnS grating was sufficiently high to provide excellent diffraction efficiency for use in the infrared wavelength region. We also present the design of a spectral disperser in CdTe for future more advanced performance.

## Introduction

An important issue in space science is the characterization of the atmospheres of exoplanets, including the challenge of detecting biomarkers. There are important molecular absorption features in exoplanetary atmospheres in the infrared wavelength region, e.g., $H_2O$ (1.13, 1.38, 1.9, 2.69, 6.2 μm), $CO_2$ (1.21, 1.57, 1.6, 2.03, 4.25 μm), $O_2$ (1.27 μm), $CH_4$ (1.65, 2.2, 2.31, 2.37, 3.3, 6.5, 7.7 μm) and $NH_3$ (1.5, 2.0, 2.25, 2.9, 3.0, 6.1, 10.5 μm), and especially $O_3$ (4.7, 9.1, 9.6 μm). Temporary differential observation of exoplanets, including the monitoring of transiting exoplanets, is one of the promising methods for characterizing the atmospheres of exoplanets. Space-borne infrared telescopes carrying a spectrometer are vital for this challenging objective. We previously presented the concept and various specialized spectrometer designs for temporary differential observations of exoplanets[1-5]. The fabrication of gratings on the surfaces of prisms made of infrared-transmissive material is a key technology for such spectrometers. ZnS is one such suitable material for a wideband infrared spectral disperser because its optical properties enable simultaneous wavelength coverage in the range of 1−13 μm[2]. In addition, it is well available and amenable for fabrication. Therefore, we carried out a trial in which we fabricated a grating on a ZnS substrate and optically evaluated it.

## Design and fabrication

The fabrication presented in this paper is the first step in our development, the primary aim of which is to study the manufacturability of gratings on ZnS. To simplify the study, we adopted a flat substrate on which to fabricate the grating rather than a ZnS prism which is more challenging. A ZnS monocrystal was used in order to realize high precision in the fabrication. Fig. 1 shows the design of the substrate and the grating. The specification is presented in Table 1.

The fabrication was executed at Canon Inc. The fabrication technology used in this work is basically the same as that described in Sukegawa et al. (2012)[6]: We utilized a high precision processing machine, the "A-former", developed in house at Canon. This machine has five axis control ($X$, $Y$, and $Z$ axes and two rotational axes) for flexible processing. To obtain high accuracy, special bearing and driving techniques have been developed, and the whole machine is installed in an environment with exceptional temperature stability. Previously, this machine had achieved a spacing accuracy of < 5 nm rms and a surface roughness of < 5 nm rms on a CdZnTe device[6].

Fig. 2 (a) shows a photograph of the fabricated ZnS sample. Fig. 2 (b) is an image of the grating taken with a scanning electron microscope, which shows that the edges of the grating are sharp. Fig. 2 (c) and (d) show the results of a measurement of microscopic 3D structure of the grating surface using a microscope interferometer, ZYGO NewView. The surface roughness obtained was 2.6 nm rms.

## Evaluation of diffraction efficiency

For our first optical testing of the ZnS sample, measurements of the diffraction efficiency were executed in air at ambient temperature using visible and near-infrared wavelength light sources. The experiments were performed in a laboratory in Koyama Astronomical Observatory in Kyoto Sangyo University. Fig. 3 shows the configurations of the experiments. First, measurement of the power at the wavelength $\lambda$ = 633 nm was done using a He-Ne laser as shown in Fig. 3 (a). A power meter was scanned in the $XY$ plane (the optical axis is in the $Z$ direction) and the position where the maximum value occurred was used in the evaluation. Because of the instability of the laser, the measurement was monitored for 1 minute. This was repeated 5 times and an average diffraction efficiency of 0.659 at $\lambda$ = 633 nm was obtained from these measurements, which are summarized in Table 2.

The diffraction efficiencies at $\lambda$ = 980 nm and 1550 nm were derived by a similar method using 980 nm and 1550 nm lasers for the light sources and a near infrared power meter of InGaAs for the measurements, as shown in Fig. 3 (b). The lasers with $\lambda$ = 980 nm and 1550 nm were more stable than the He-Ne laser, so the output could be used directly (i.e., without the need to monitor the values for 1 minute). The results of averaging 5 measurements gave diffraction efficiencies of 0.674 and 0.702 at $\lambda$ = 980 nm and 1550 nm, respectively (Table 2).

The diffraction efficiency was also estimated by calculations using a Fourier Modal Method (FMM) as summarized in Table 2. The refractive index of ZnS was obtained from the literature[7]. For the incident side, the Fresnel transmittance with the angle of incidence perpendicular to the surface was calculated. For the side through which light exits, the light diffracted by the grating was calculated using FMM. Then the efficiency of the diffraction order providing the maximum power was determined. Finally the diffraction efficiency was derived from the efficiencies of the two surfaces for $\lambda$ = 633 nm, 980 nm, and 1550 nm (Tables 1 and 2).

The results show that the difference between the diffraction efficiencies obtained by experiment and by calculation were between 0.9 and 2.4 %. We concluded that the quality of the fabrication of the ZnS sample was good enough to provide excellent diffraction efficiency for applications in the infrared wavelength region.

## Discussion: future work in development and application to an astronomical spectrometer

Since the fabrication of a grating on a flat substrate was successful, the next step is to develop a prism with a grating surface to realize a wideband spectral disperser. Important future work for us is to pursue a laboratory demonstration and detailed evaluation of a spectrometer with full optics in a wide infrared wavelength region at cryogenic temperatures using the prototype-testbed for Infrared Optics and Coronagraph (PINOCO) in ISAS[8].

Optimization of the design of the spectrometer is also important. CdTe and CdZnTe are harmful materials but they enable optically wider coverage, $\lambda = 1-33$ μm[1]. A Si:As detector can cover most of this wavelength region and it makes the instrument very simple while with a combination of detectors (e.g., a Si:As detector and an InSb detector) higher efficiency can be realized[2]. Fig. 4 (a) and (b) present an example design of a spectrometer using CdTe and its spectral format, respectively. It should be noted that the extreme aberration is intentionally given in the design shown in Fig. 4. (Hereafter we refer to this aberration as "defocusing"). As a result of the defocusing, the point spread function (PSF) is diffused and the peak of the PSF is suppressed. The primary aim of defocusing is to improve the observational stability which is a key issue for characterizing exoplanetary atmospheres. Another important aim is to improve the exposure efficiency, [exposure time]/[exposure+readout time], drastically by relaxing the saturation problem. In the design of Fig. 4 (b), for instance, we assumed a 3 m telescope with a diffraction limited wavefront error of $\lambda = 20$ μm, and defocusing corresponding to a diffraction limited PSF size of $\lambda = 20$ μm and $\lambda = 40$ μm for *X* and *Y* direction at the geometric optics approximation, respectively. It should be noted that a 1 m class telescope has the potential to gather photons to detect $O_3$ features in exoplanetary atmospheres[9, 10]. So a 3 m class telescope can do it more efficiently. Fig. 5 (a) presents the results of simulations of the observational stability assuming these optics and the following conditions; the pixel scale of the detector is Nyquist sampling at $\lambda = 10$ μm, and the pixel-to-pixel inhomogeneity of the sensitivity of the array detector is 1 % (1σ, random Gaussian distribution). To simulate the inhomogeneity of the sensitivity in each pixel, the pixels are each divided into 3×3 square areas and 10 % higher sensitivity is given to the central area than the other 8 areas. Fig. 5 (a) shows that the influence of the telescope pointing error on the stability is drastically reduced by defocusing. Furthermore, averaging the effect of the telescope pointing jitter through exposure reduces the instability of the data. As a result, the simulated stability reaches the required value for the characterization of exoplanets ($10^{-4}$ as nominal, and $10^{-5}$ as the goal)[9, 10] with following conditions; telescope pointing stability of 0.07 arcsec (1σ, random Gaussian distribution), pointing jitter frequency of 15 Hz, and exposure time of 3600 sec. Another benefit of defocusing is that the wavelength resolving power, *R*, remains suitable for studying the molecular features in exoplanetary atmospheres over a very wide wavelength region as shown in Fig. 4 (b). Furthermore, the slit width of the spectrometer with defocusing can be adjusted for a very wide wavelength region as shown in Fig. 5 (b). It is interesting to note that dedicated exoplanet science is possible using a satellite borne 3 m infrared telescope with specifications limited to $\lambda = 20$ μm, at very low additional cost, and with technically robust instrumentation in which neither active wavefront correction nor moving parts are required.

## Acknowledgement

This work is supported by a fund of JAXA, Rijicho-sairyo-keihi. We are grateful for Keiji Tachikawa's approval to allow us flexible use of this fund.

Table 1. Specifications.

| Material | ZnS monocrystal |
|---|---|
| Geometry of substrate | L: 10 mm, W: 10 mm, t: 1 mm at the thickest part. One surface is with grating and the other surface is flat. (See Fig. 1 for more detail). |
| Blaze angle | 2.1 ± 0.1 deg. |
| Pitch of grating | 166.667 ± 0.015 (priodic) ± 0.01 (ramdom) μm |
| Sharpness of grating edge | R < 0.1 μm |
| Form torelance | < 0.5 μm PV |
| Surface roughness | < 10 nm rms |
| Imperfectoins | 40−60  cratch-dig |

Table 2. Summary of the optical evaluation of the diffraction efficiency.

| | | $\lambda$ = 633 nm | | | | | | |
|---|---|---|---|---|---|---|---|---|
| | Meas. | Power [mW] (with ZnS sample) | | | Power [mW] (no ZnS sample) | | | Diffraction efficiency |
| | | max | Min | Ave. | max | min | Ave. | |
| Meas. Data | 1 | 1.109 | 1.033 | 1.071 | 1.660 | 1.549 | 1.605 | 0.667 |
| | 2 | 1.104 | 1.029 | 1.067 | 1.696 | 1.575 | 1.636 | 0.652 |
| | 3 | 1.102 | 1.025 | 1.064 | 1.701 | 1.581 | 1.641 | 0.648 |
| | 4 | 1.125 | 1.048 | 1.087 | 1.698 | 1.583 | 1.641 | 0.662 |
| | 5 | 1.133 | 1.048 | 1.091 | 1.700 | 1.585 | 1.643 | 0.664 |
| | Ave. | | | | | | | **0.659** |
| FMM | | | | | | | | **0.683** |

| | | $\lambda$ = 980 nm | | | $\lambda$ = 1550 nm | | |
|---|---|---|---|---|---|---|---|
| | Meas. | Power [mW] (with ZnS sample) | Power [mW] (no ZnS sample) | Diffraction efficiency | Power [mW] (with ZnS sample) | Power [mW] (no ZnS sample) | Diffraction efficiency |
| Meas. data | 1 | 2.66 | 3.95 | 0.673 | 1.419 | 2.034 | 0.698 |
| | 2 | 2.65 | 3.93 | 0.674 | 1.428 | 2.027 | 0.704 |
| | 3 | 2.63 | 3.91 | 0.673 | 1.431 | 2.030 | 0.705 |
| | 4 | 2.64 | 3.91 | 0.675 | 1.430 | 2.033 | 0.703 |
| | 5 | 2.64 | 3.91 | 0.675 | 1.425 | 2.036 | 0.700 |
| | Ave. | | | **0.674** | | | **0.702** |
| FMM | | | | **0.697** | | | **0.711** |

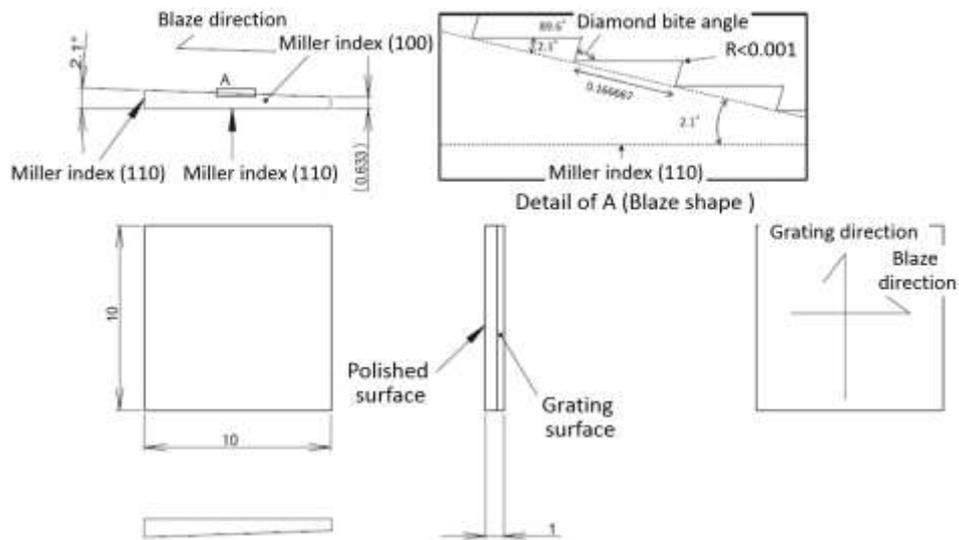

Fig.1. Design of the ZnS substrate and the grating on the surface of the substrate.

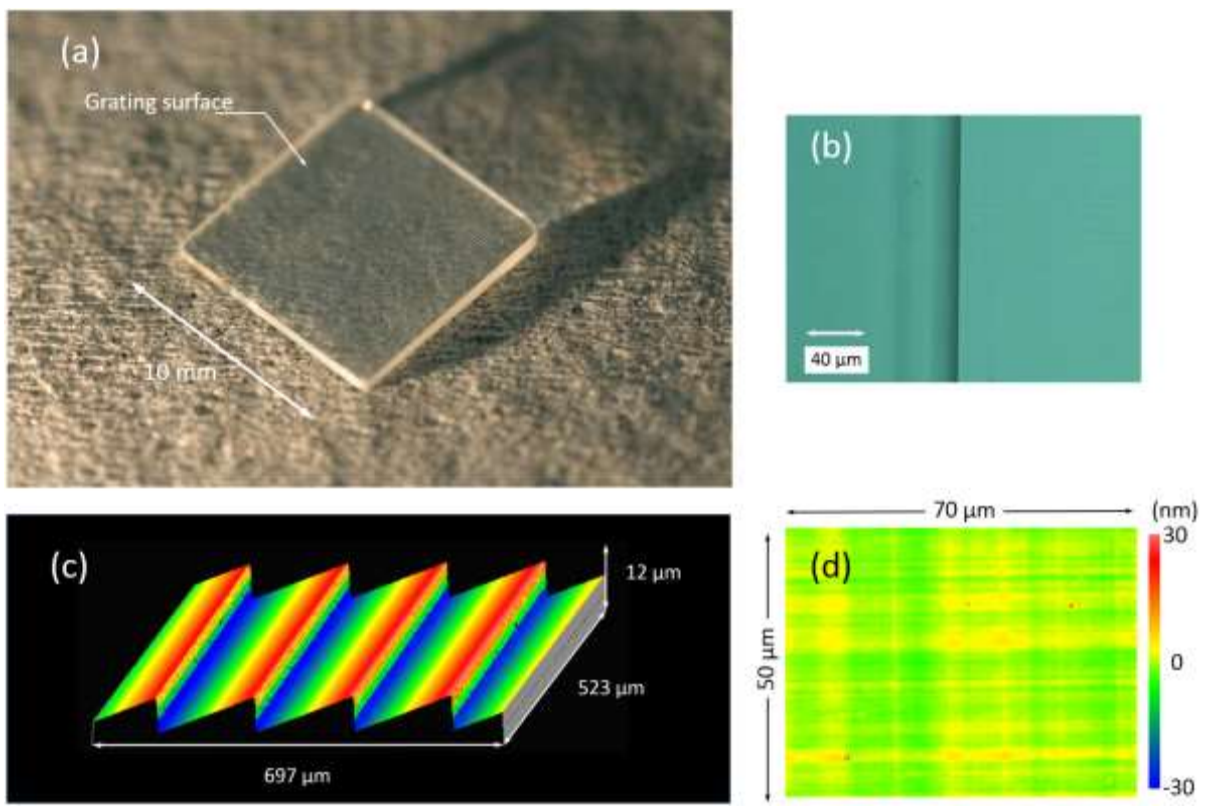

Fig. 2. (a): Fabricated ZnS sample. (b): An image of the grating edge taken with a scanning electron microscope. (c), (d): Microscopic 3D structure of the grating surface obtained with a microscope interferometer, ZYGO NewView.

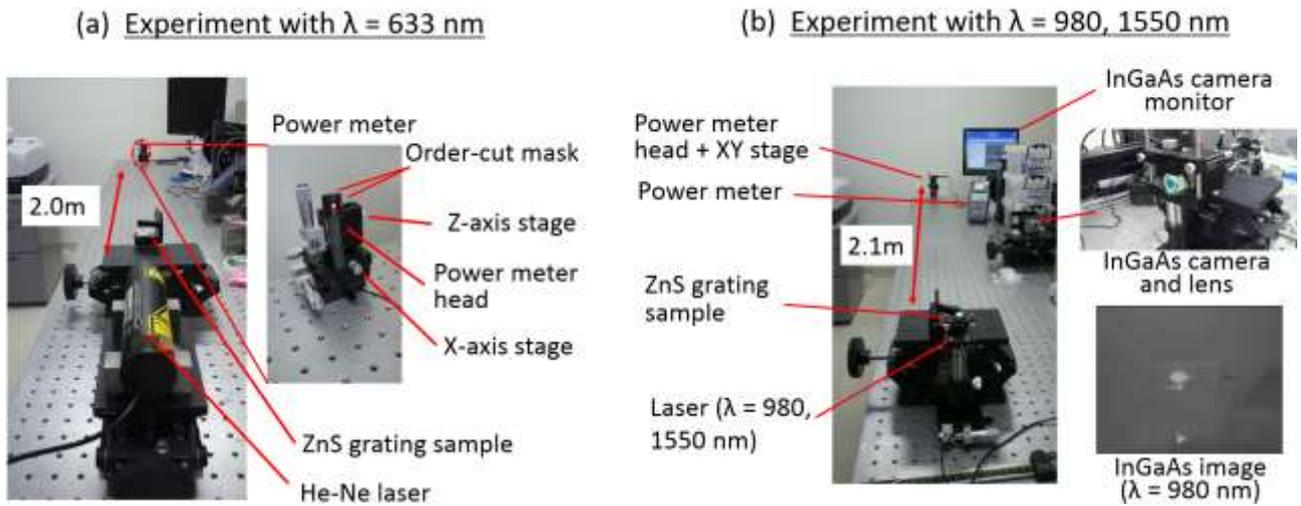

Fig. 3. (a): Configuration for evaluating the diffraction efficiency at λ = 633 nm. (b): Configuration for evaluating the diffraction efficiency at λ = 980 nm and 1550 nm.

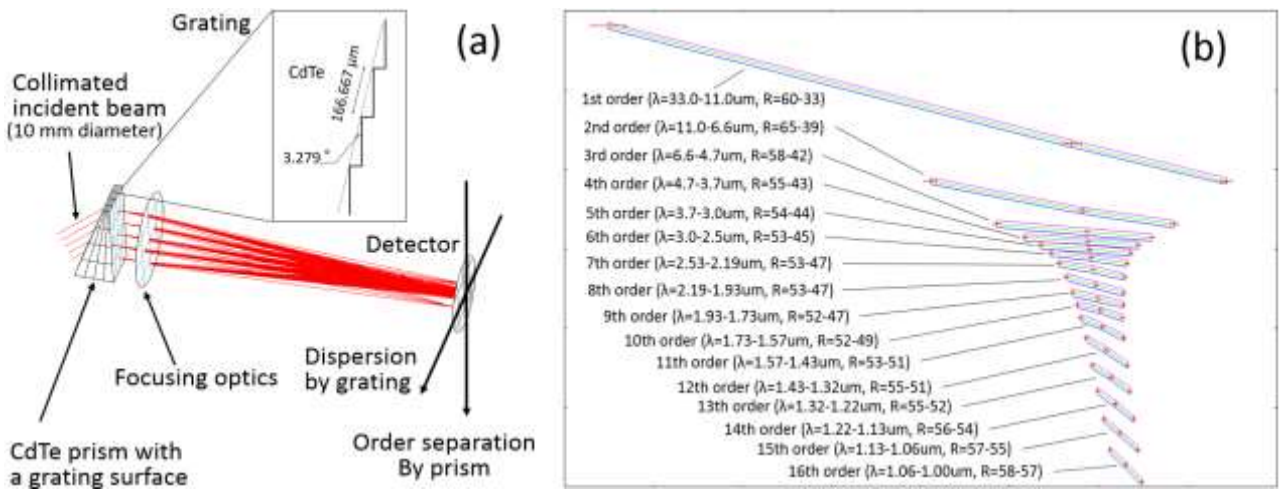

Fig. 4: (a): Design of a wideband infrared spectrometer with a spectral disperser made of CdTe. (b): spectral format of the spectrometer.

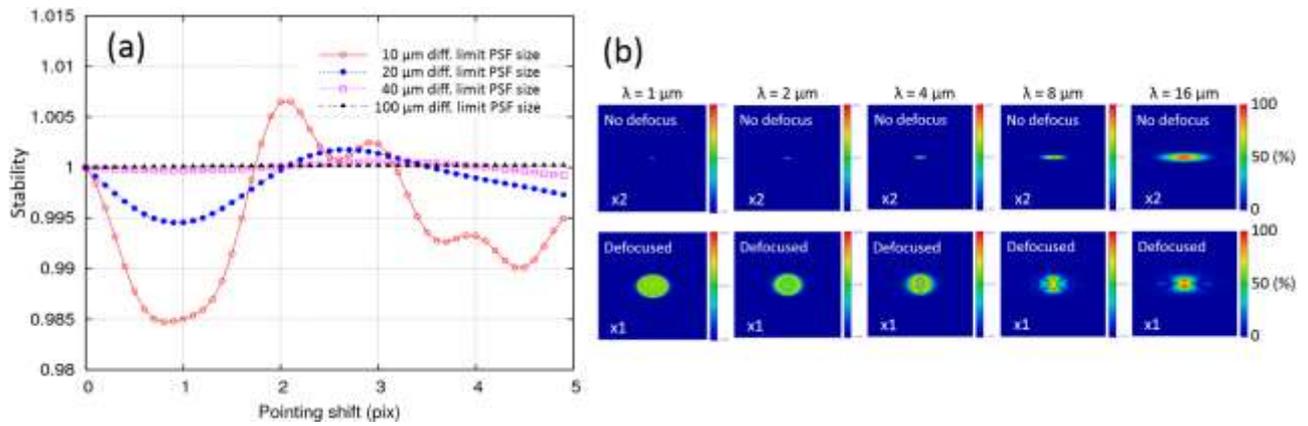

Fig. 5. (a): Results of simulation for influence of defocusing on data stability. (b): Spot diagrams for optics with and without defocusing.